\documentclass[12pt]{article}
\usepackage{amssymb}
\usepackage{amsmath}
\usepackage{amsfonts}
\usepackage{graphicx}
\usepackage{mathrsfs}
\usepackage{comment}

\newcommand{\Slash}[1]{{\ooalign{\hfil#1\hfil\crcr\raise.167ex\hbox{/}}}}

\newcommand{\beq}{\begin{equation}}  \newcommand{\eeq}{\end{equation}}
\newcommand{\bef}{\begin{figure}}  \newcommand{\eef}{\end{figure}}
\newcommand{\bec}{\begin{center}}  \newcommand{\eec}{\end{center}}
\newcommand{\non}{\nonumber}  \newcommand{\eqn}[1]{\beq {#1}\eeq}
\newcommand{\laq}[1]{\label{eq:#1}}  

\newcommand{\Eq}[1]{Eq.(\ref{eq:#1})}
\newcommand{\Sec}[1]{Sec.\ref{chap:#1}}
\newcommand{\eq}[1]{(\ref{eq:#1})}
\newcommand{\ab}[1]{\left|{#1}\right|}

\def\({\left(}
\def\){\right)}

\def\O{\mathcal{O}}

\def\ebq{\eeq \beq}

\newcommand{\GEV}{ {\rm GeV} }
\newcommand{\TEV}{ {\rm TeV} }
\def\o{\over}

\def\b{\beta}

\def\d{\delta}

\def\f{\phi}

\def\m{\mu}

\def\p{\psi}

\def\s{\sigma}
\def\t{\tau}

\def\tl{\tilde}
\def\*{\dagger}

\begin{document}

\begin{titlepage}
\begin{center}

\hfill TU-1000\\

\vspace{2.5cm}

{\LARGE \bf 
A Novel Approach to Fine-Tuned \\
Supersymmetric Standard Models\\}

\vspace{0.5cm}
{\large \bf --Case of Non-Universal Higgs Masses model--}
\vspace{2.5cm}

 {\bf Masahiro Yamaguchi\footnote{e-mail: yama@tuhep.phys.tohoku.ac.jp}} and  {\bf  Wen Yin\footnote{email: yinwen@tuhep.phys.tohoku.ac.jp}}

\vspace{1.5cm}
{\it Department of Physics, Tohoku University, Sendai 980-8578, Japan} 
\vspace{2.0cm}

\begin{abstract}
Discarding the prejudice about fine tuning, we propose a novel and efficient approach
 to identify relevant regions of fundamental parameter space in supersymmetric models with some amount of fine tuning.
The essential idea is the mapping of experimental constraints at a low energy scale, rather than the parameter sets, 
to those of the fundamental parameter space. Applying this method to the non-universal 
Higgs masses model, we identify a new interesting superparticle mass pattern where some of the first two generation
 squarks are light whilst the stops are kept heavy as 6TeV. Furthermore, as another application of this method, we show that the discrepancy
  of the muon anomalous magnetic dipole moment can be filled by a supersymmetric contribution
   within the $1~\s$ level of the experimental and theoretical errors,
   which was overlooked by the previous studies due to the required terrible fine tuning.

\end{abstract}

\end{center}
\end{titlepage}
\setcounter{footnote}{0}

\section{Introduction}

Although the discovery of the Higgs boson in July 2012 verifies our thought that the physics up 
to the electroweak scale should be well described by the standard model (SM) of particle physics
 \cite{Aad:2012tfa}, the SM itself suffers from the uncomfortably large disparity between the electroweak
  scale and the fundamental physics scale which is supposedly close to the Planck scale. 
Supersymmetry (SUSY) has been recognized as a promising candidate to solve this unease. The fact that
 superparticles have not yet been discovered, however, constrains their mass spectra, if exists: e.g. 
 colored superparticles should weigh at least around 1 TeV \cite{dEnterriaCMS:2015jva}.  In the 
 minimal supersymmetric standard model (MSSM), the measured Higgs boson mass of about 125 GeV
  requires large radiative corrections due to supersymmetry breaking (SUSY-breaking) to raise its tree-level mass below the 
  Z boson mass  \cite{Okada:1990vk}. As the Higgs boson strongly couples to the top-stop sector, this typically
   requires that the stop mass has to be around 6 TeV or so, unless a SUSY-breaking trilinear coupling
    is parametrically large \cite{Hahn:2013ria}. In this case, there will be a little hierarchy between the electroweak scale and
     SUSY-breaking mass parameters, and thus some amount of fine tuning among these parameters may
      be requisite in order that the electroweak symmetry breakdown takes place at the correct energy scale.

 This situation does not mean that nature rejects SUSY, but implies that we should not have prejudice
  against the amount of fine tuning. Since SUSY is still a promising candidate for physics beyond the
   SM, we should study the supersymmetric SM with some amount of fine tuning (FT-SUSY: fine-tuned supersymmetry).

To identify an experimentally viable region or an interesting region of a model, 
the scatter plot method has been widely used. This method represents a relevant region
 by a collection of discretized points in the fundamental parameter space, just like a ``pointillism".
 The collection of points is selected from a large number of initially chosen points in the fundamental parameter space to satisfy
 the experimental (and other) constraints at the experimental scale. 
 However, in a FT-SUSY the relevant region might be too tiny to be represented in this way.\\

In this paper, we propose a novel approach to a FT-SUSY regardless of the amount of fine tuning.

In \Sec{approach}, we propose a method to identify the relevant region of a FT-SUSY.
 In contrast to the ordinary top-down renormalization group (RG) picture, in which a point chosen in the fundamental parameter space at the fundamental scale flows to that at the experimental scale, 
 we map a constraint for the parameter space at the experimental scale to that at the 
 fundamental scale. Then, we can directly identify the restricted space by the mapped constraints as the relevant region written in 
 the fundamental parameters. 
This procedure is like a ``coloring". 
This procedure allows us to identify the whole relevant region as well as its outlines in the
 fundamental parameter space. Furthermore, the area near an outline can be easily identified
  as a phenomenologically interesting region, if this outline corresponds to the boundary
   of a constraint given by an on-going experiment.  
Since the constraints we map can also include the requirement of
 a characteristic property, if we choose a suitable requirement, a fine-tuned region is identified. 

In \Sec{NUHM}, to illustrate our idea and to show its efficiency, we apply this procedure to the non-universal Higgs masses
 model (NUHM) \cite{Berezinsky:1995cj} which has the 
MSSM particle contents with universal SUSY breaking masses except for the Higgs masses at the GUT scale. We identify the experimental viable region of the NUHM and argue its features. 
We find an interesting region with a new superparticle mass pattern, where some of the first two 
generation squarks are light (\Sec{NUHMregion}).
This mass pattern is a consequence of the RG running where a negative Higgs mass squared dominantly raises the third generation squark masses due to their rather large Yukawa couplings. 
 Since this effect never happens in the CMSSM, this region should be one of the characters of the NUHM (\Sec{ILSQ}).

In \Sec{muong-2}, using another application of our method, we find there is a terribly fine-tuned region that explains the anomaly of the muon
 anomalous magnetic dipole moment (muon $g-2$)  \cite{Bennett:2006fi}, \cite{Hagiwara:2011af}, \cite{Davier} within the $1~\s$ experimental and theoretical errors
in the NUHM. Furthermore, with sufficiently large $\tan\beta$, we also show that there is a parameter region
   explaining the muon $g-2$ anomaly with most of the 1st and 2nd generation sfermions light. 
 These regions were overlooked by the previous studies using the scatter plot method which is not practical to find such a tiny and terribly fine-tuned region. This fact shows the power of our approach to a FT-SUSY.

\section{A Novel Approach to FT-SUSY}

We propose a novel approach to tackle a FT-SUSY.\footnote{In fact, the method developed here can apply to many models even without SUSY. However, for ease of explanation, we only apply our method to a FT-SUSY in this paper.}
In this approach, we can directly identify the relevant region in the FT-SUSY without being bothered with some amount of fine tuning.

\label{chap:approach}

For the sake of simplicity, suppose that a fundamental supersymmetric model, such as a grand unified 
theory (GUT), can be described as a generic MSSM. The generic MSSM is defined as an effective theory 
with most general SUSY-breaking soft mass parameters of the particle contents, below the fundamental scale 
$t_f$ of the fundamental model. $t_f$ could be $\log{\left( 10^{{\O (10)}}{\rm GeV} \o m_z \right)}$ 
depending on the model we consider, where $m_z$ is the Z boson mass.  
 In the parameter space of the generic MSSM, $\mathcal{M}_f^{\rm gen}$, a point is specified by a set of $
 \O(100)$ dimensionful parameters, $g^f_i$ at the scale $t_f$, where we have assumed $\tan\beta$ is a 
 given constant and no parameters are dimensionless. This assumption is only for simplicity and the 
 generalization is straightforward.
In contrast the fundamental model has a restricted parameter space, the fundamental parameter space $\mathcal{M}_f^{\rm fund}$, with coordinates of much fewer fundamental parameters $G_a$.  Since below the scale $t_f$, 
the fundamental model is described by the generic MSSM, $\mathcal{M}_f^{\rm fund}$ is embedded into a subspace of $\mathcal{M}_f^{\rm gen}$ by a set of relations,  
\beq
g_i^f=f_i^f(G_a).
\eeq
This defines a map,
\beq
\laq{sub}
f^f:  \mathcal{M}_f^{\rm fund} \rightarrow \mathcal{M}_f^{\rm gen}, ~{\rm and}~  f^f(\mathcal{M}_f^{\rm fund})=\left\{ g^f_i \in \mathcal{M}_f^{\rm gen} | g^f_i=f^f_i(G_a),~ G_a \in \mathcal{M}_f^{\rm fund} \right\}.
\eeq

On the other hand, the solution of the RG equation \cite{Martin:1993zk} which gives correspondence among the parameters of the same theory at different scales can also be considered as a map $f_{RG}$.\footnote{For simplicity, we suppose that $f_{RG}$ is a bijection, so that the image satisfies the equality, ${\rm im}f_{RG}\equiv f_{RG}(\mathcal{M}_f^{\rm gen})=\mathcal{M}_e^{\rm gen}$, and the inverse map, $f^{-1}_{RG}$, can be defined.}Since the fundamental model can be described by the generic MSSM, we consider $f_{RG}$ in the context of the generic MSSM:
\beq
\laq{RG}
f_{RG}:  \mathcal{M}_f^{\rm gen} \rightarrow \mathcal{M}_e^{\rm gen},~{\rm and}~  f_{RG}(\mathcal{M}_f^{\rm gen})=\left\{ g^e_i \in \mathcal{M}_e^{\rm gen} | g^e_i=g^{sol}_i(t_e; t_f, g^f_j),~ g_i^f \in \mathcal{M}^{\rm gen}_f \right\}. 
\eeq 
$\mathcal{M}_e^{\rm gen}$ is the parameter space at the experimental scale $t_e=\log{({\O (100)\GEV \o m_z})}$, and a set of $g^{sol}_i(t_e; t_f, g^f_j)$ is the solution of the RG equation \cite{Martin:1993zk} in the generic MSSM at $t_e$ with an initial condition of a set of parameters, $g_j^f$ at $t_f$.

Suppose that $\tl{\mathcal{M}}_{e}^{\rm gen}$ denotes the region of interest of the generic MSSM at $t_e$, which may be either a viable region, i.e. the part of the parameter space that survives the experimental constraints, or a phenomenologically interesting region with some characteristic properties.
$\tl{\mathcal{M}}_e^{\rm gen}$ is characterized by a set of conditions expressed 
as $\f_l(g^e_i)>0$ or $\f_l(g^e_i)=0$:
\beq
\laq{EC}
\mathcal{M}_e^{\rm gen} \supset \tl{\mathcal{M}}_e^{\rm gen}=\left\{ g^e_i \in \mathcal{M}_e^{\rm gen} | \f_1(g^e_i)>0, \f_2(g^e_i)>0,.... , \f_n(g^e_i)>0 \right\}.
\eeq
Here $\f_l(g^e_i)$ is a condition function for the generic MSSM parameters at $t_e$, which could either correspond to a fitted function of an experimental constraint or a requirement to have a characteristic property. The conditions in equalities, such as the ones for correct electroweak symmetry breaking and the Higgs boson mass, reduce the dimension of $\mathcal{M}_e^{\rm gen}$.  On the other hand, the conditions in inequalities, such as the mass bounds for superparticles, restrict the parameter space $\mathcal{M}_e^{\rm gen}$ and hence constitute the outlines of $\tl{\mathcal{M}}_e^{\rm gen}$. In \Eq{EC}, we have written down only the conditions in inequalities for illustrative purpose.

What we would like to do is to identify the region of interest at $t_f$, $\tl{\mathcal{M}}_f^{\rm fund}$ in the parameter space of the fundamental model, $\mathcal{M}_f^{\rm fund}$.
A conventional definition of $\tl{\mathcal{M}}_f^{\rm fund}$ is given as  
\beq
\laq{viable}
\mathcal{M}_f^{\rm fund}\supset \tl{\mathcal{M}}_f^{\rm fund}=\left\{ G_a \in \mathcal{M}_f^{\rm fund} | {f_{RG}} \circ f^{f}(G_a) \in \tl{\mathcal{M}}_e^{\rm gen} \right\}.
\eeq
Namely, given a set of the fundamental parameters, $G_a \in \mathcal{M}_f^{\rm fund}$, we apply the RG procedure to obtain the corresponding parameters at $t_e$, and check whether they satisfy the conditions characterizing the region of interest of the generic MSSM $\tl{\mathcal{M}}^{\rm gen}_e$.

The ordinary scatter plot method follows this procedure recursively by using sample points, $S={\left\{ G^{(1)}_a,  G^{(2)}_a, ... G^{(N)}_a \right\}}$, which are chosen in some way from the fundamental parameter space. Here $N$ is the total number of the sample points. The region of interest of the fundamental model, $\tl{\mathcal{M}}^{\rm fund}_f$ is approximated as a collection of discretized points, like a ``pointillism". 
Therefore, when applying to a FT-SUSY, in which the region of interest is so tiny, the ordinary method requires a huge number of sample points, $N$, as well as luck, and hence is time-consuming in numerical computation and sometimes misleading.
\\

We now propose a novel approach to identify the region of interest, $\tl{\mathcal{M}}_f^{\rm fund}$, in a fundamental model regardless of the amount of fine tuning $\tl{\mathcal{M}}_f^{\rm fund}$ has. 
Our definition of $\tl{\mathcal{M}}_f^{\rm fund}$ can be written,
\beq
\laq{appro}
\tl{\mathcal{M}}_f^{\rm fund}=\left\{ G_a \in \mathcal{M}_f^{\rm fund} | \psi_1(G_a)>0, \psi_2(G_a)>0,.... , \psi_n(G_a)>0 \right\},
\eeq
where $\psi_l(G_a)$ is a condition function for the fundamental parameter space,
expressed as
\beq
\laq{fundcon}
\psi_l(G_a)=\f_l \left(g_i^{sol}\left(t_e; t_f, f^f_j(G_a)\right)\right)=  \f_l \circ f_{RG}\circ f^f(G_a),
\eeq
and hence should be equivalent to \Eq{viable}. However, what we would like to obtain is not the correspondence among points in $\mathcal{M}^{\rm fund}_f$ and $\tl{\mathcal{M}}^{\rm gen}_e$, but the correspondence between the two condition functions, $\f_l(g_i^f)$ and $\p_l(G_a)$.    
Namely, we map the given set of conditions, $\f_l(g_i^e)>0$ that characterizes $\tl{\mathcal{M}}_e^{\rm gen}$, to the corresponding one, $\f_l \circ f_{RG}(g_i^f)>0$, for the parameter space $\mathcal{M}_f^{\rm gen}$ at $t_f$ within the generic MSSM, and transform the latter into the corresponding conditions in $\mathcal{M}^{\rm fund}_f$. Since the boundaries of these mapped conditions constitute the outlines of $\mathcal{M}^{\rm fund}_f$ and what we identify as $\tl{\mathcal{M}}_f^{\rm fund}$ is the interior of the outlines, our procedure is like a ``coloring".\\

 Since the map, $f^f$, is given, what we would like to know is the RG map of the condition function, $\f_l(g_i^e)$, within the generic MSSM,
\beq
\laq{mapcon}
\f_l \circ f_{RG}(g_j^f)=\f_l \left(g_i^{sol}(t_e; t_f, g_j^f)\right),
\eeq
 and we will show how to derive the explicit form of this. One way is to solve the RG equation of the generic MSSM so that we can express $g_i^{sol}(t_e;t_f,g_j^f)$ in terms of the set of the parameters ${g}_j^f$ at $t_f$.

Alternatively, we can solve the differential equation which follows the RG map of the condition function, $\Phi_l(g_j,t) \equiv \f_l \left(g_i^{sol}(t_e; t, g_j)\right)$,
by varying $t$:
\begin{align}
\label{eq:gfRECC}
\biggl(\frac{\partial}{\partial t}+\sum_i \beta _i \frac{\partial}{\partial g_i}\biggr)\Phi_l(g_j,t)=0,
\end{align}
where $\b_i$ is the RG beta function for $g_i$ \cite{Martin:1993zk}.

If the perturbative expansion
\begin{align}
\laq{tay}
\Phi_l (m_j,t)=\sum_{n=0}{1 \o n!} \phi^{i_1,i_2...i_n}_l(t)g_{i_1}g_{i_2}...g_{i_n}
\end{align}
is allowed, a set of linear differential equations  
\begin{align}
\laq{RECC}
\frac{\partial}{\partial t}\phi^{i_1,i_2...i_n}_l(t)&= \sum_{m=1}^{n} \tilde{\beta}^{i_1,i_2...i_n}_{j_1,j_2...j_m}
\phi^{j_1,j_2...j_m}_l(t),
\end{align} for the coefficients, $\phi^{i_1,i_2...i_n}(t),$ are derived by requiring the vanishing of each Taylor 
coefficient in \Eq{gfRECC}. The upper limit of the summation in \Eq{RECC} comes from the fact that 
a perturbative RG  beta function always contains parameters of total exponents $\geqslant 1$. \Eq{RECC} is the 
running equation for coefficients of constraint (RECC) for $\f_l(g^e_i)$. If all the parameters are 
dimensionful as in our case, $ \tilde{\beta}^{i_1,i_2...i_n}_{j_1,j_2...j_m}$ in \Eq{RECC} can be non-zero only if
the dimension of $g_i$, $d(i)$, satisfies $\sum_{l}^n{d(i_l)}=\sum_{l}^m{d(j_l)}$. Hence the coefficients can be 
evaluated by numerically solving the derived linear differential equation.\footnote{If dimensionless parameters 
are included, we can also solve \Eq{RECC} but only perturbatively, namely the solution can approximately 
represent a mapped constraint function up to a precision depending on the order of the couplings we take into 
account.}
We show the explicit derivation of RECC in the generic MSSM in Appendix \ref{chap:MSSM}.

Notice that $g^{sol}_i(t_e; t_f, g^f_i)$ in terms of a set of parameters, $g^f_i$, is obtained, once we choose a condition function $\f_i(g^e_j)=g^e_i$. 
We also note that in the derivation of RECC, $\beta_i$ and $\tilde{\beta}^{i_1,i_2...i_n}_{j_1,j_2...j_m}$ can even depend on the scale $t$. This is a convenient fact because we may take a shortcut to derive RECC with some parameters approximately treated as constants. Namely, if possible, we can numerically solve the RG equations for these parameters in advance, and substitute the numerical solutions as constants in the remaining RG equations. Then we can derive RECC from these remaining RG equations which explicitly depend on $t$. This is what we do in Appendix \ref{chap:MSSM}. 

Solving the corresponding RECCs, we can obtain the set of conditions in terms of $g^f_i$, $\f_l\circ f_{RG}(g^f_i)>0$, and applying the given map, $f^f$, a set of $\p_l(G_a)>0$ is derived from \Eq{fundcon}.
Therefore the whole region of interest, $\tl{\mathcal{M}}^{\rm fund}_f$, is identified from \Eq{appro}.
\\

There are two additional advantages in our approach. 

Since a boundary of a constraint could correspond to an outline of $\tl{\mathcal{M}}^{\rm fund}_f$, the viable region near such an outline may be testable if this constraint is given by an ongoing experiment. This implies that a viable region near an outline can be a phenomenologically interesting region. Therefore, checking the boundary profiles, namely the constraints the boundaries correspond to, we can guess some of the phenomenological interesting regions even without any additional requirement of characteristic properties.
On the other hand, the boundary profiles of a phenomenologically interesting region in turn suggest the predictions that can be accompanied with the characteristic property.

The second advantage is due to the fact that in our procedure the RG map of a condition function is followed within the MSSM. In fact, we can define the region of interest of the generic MSSM at $t_f$, 
\beq
\tl{\mathcal{M}}_f^{\rm gen}=\left\{ g_i^f \in \mathcal{M}_f^{\rm gen} | \f_1\circ f_{RG}(g^f_i)>0, \f_2\circ f_{RG}(g^f_i)>0,.... , \f_n\circ f_{RG}(g^f_i)>0 \right\}.
\eeq
Since the viable region of the generic MSSM at $t_f$ directly responds to the fundamental model, with stringent enough experimental constraints in future, $\tl{\mathcal{M}}_f^{\rm gen}$ can be a probe of the fundamental model.  This approach may clarify the fundamental model directly.

\section{Region of Interest in the Non-Universal Higgs Masses Model}
\label{chap:NUHM}
Using the method advocated in the previous section, we would like to analyze the non-universal-Higgs 
masses model (NUHM)\cite{Berezinsky:1995cj} as an example of a fundamental model. The NUHM is an 
extension of the CMSSM motivated by GUT and has universal masses for 
sfermions at the GUT scale $t_f\sim \log{ \left( 2 \times 10^{16} \GEV \o m_z\right)}$. The only difference from 
the CMSSM is that in the NUHM the SUSY breaking Higgs mass squared parameters are allowed to vary from that of sfermions at the 
GUT scale. This may be a natural assumption as the origin of the 
Higgs particles may be different from those of sfermions.

The NUHM has a fundamental parameter space, $\mathcal{M}_f^{\rm fund}$, where a point is specified by the 
fundamental parameters,
\eqn{\laq{fp} G_a= \left\{ m_0^2,m_{\rm Hu0}^2,m_{\rm Hd0}^2, M_0,A_0, \mu_0,B_0\right\}.} 
 As we have noted, $\tan\beta$, as well as the other dimensionless couplings, is taken to be a given constant that is not included in the parameter set, \Eq{fp}. 
Since the NUHM can be described by the generic MSSM below $t_f$, 
a set of fundamental parameters in $\mathcal{M}_f^{\rm fund}$ is related to the parameters in $\mathcal{M}_f^{\rm gen}$: 
\beq
\laq{unis}
{\bf m}_{\rm \tl{Q}}^2={\bf m}_{\rm \tl{u}}^2={\bf m}_{\rm \tl{d}}^2={\bf m}_{\rm \tl{L}}^2={\bf m}_{\rm \tl{e}}^2=m_0^2 {\bf 1}
\ebq
\laq{unig}
M_{\rm 1}=M_{\rm 2}=M_{\rm 3}=M_0
\ebq
\laq{univA}
{\bf A}_u={\bf A}_d={\bf A}_e=A_0 {\bf 1}
\ebq
\laq{nonuniv}
m_{\rm Hu}^2=m_{\rm Hu0}^2,~ m_{\rm Hd}^2=m_{\rm Hd0}^2
\ebq
\laq{B}
B=B_0, \m=\m_0.
\eeq 
Eqs. \eq{unis}, \eq{unig} and \eq{univA} are the conditions of universal sfermion mass, gaugino mass and $A-$term, respectively, where the bold characters are understood as three by three matrices of generation. \Eq{nonuniv} expresses the condition of the non-universal Higgs masses at $t_f$ which is the only difference from the CMSSM, and also is the character of this fundamental model.  The Higgs mixing parameter, $\mu$, and the $B-$ term are taken to be free at $t_f$ in \Eq{B}. 

The experimental constraints of the generic MSSM at the experimental scale, $t_e \sim \log({100\GEV \o m_z})$, that restrict $\mathcal{M}_e^{\rm gen}$ to its viable region $\tl{\mathcal{M}}_e^{\rm gen}$ are given:
\begin{align}
\label{eq:MaC1}
\begin{split}
&2B\mu -(m^2_{{\rm {\rm Hu}}}+m^2_{{\rm Hd}}+2\mu ^2)\sin 2\beta=0\\
&\mu ^2-\frac{(m^2_{{\rm Hd}}-m^2_{{\rm Hu}}\tan^2{\beta} )}{\tan^2{\beta} -1}-\frac{m^2_{z}}{2}=0
\end{split}\\
\label{eq:MaC2}
&m^2_{{\rm \tilde{Q}3}}\cdot m^2_{{\rm \tilde{u}3}} \equiv m_{{\rm soft}}^4 \sim (6{\rm TeV})^4\\
\label{eq:MaC3}
\begin{split}
&m^2_{{\rm \tilde{Q}}i,{\rm \tilde{u}}i,{\rm \tilde{d}}i} > (1 {\rm TeV})^2,m^2_{{\rm \tilde{L}}i,{\rm \tilde{e}}i}>(300{\rm GeV})^2 \\
&M_3^2>(2 {\rm TeV})^2, \mu^2> (300{\rm GeV})^2,m_{\rm A}^2 >(300 {\rm GeV})^2. 
\end{split}
\end{align}
Here Eqs.(\ref{eq:MaC1}) are the constraints to obtain a correct electroweak vacuum at the tree level.
Eq.(\ref{eq:MaC2}) is a rough requirement for the SM Higgs boson mass around $\sim 125$GeV suggested by FeynHiggs 2.11.2 \cite{Hahn:2013ria}.
Eqs.(\ref{eq:MaC3}) are the LHC and LEP bounds of the superparticles in the MSSM \cite{dEnterriaCMS:2015jva}.
Neglecting the 1st and 2nd generation Yukawa couplings, the SU(2) flavor symmetry suppresses flavor violation in the sfermion sector,
 and we do not consider the constraints from flavor physics. Also we assume that the parameters \Eq{fp} are real and do not consider constraints of CP violation. Since we will solve the RECCs at the 1loop level, which can be derived from the given 1loop RG equations of the MSSM as in Appendix \ref{chap:MSSM}, we ignore the threshold corrections to the parameters. To apply a higher loop analysis, we can include the threshold corrections in these constraints and solve the RECCs derived from the higher loop RG equations.

\subsection{Whole Viable Region and Phenomenologically Interesting Regions}
\label{chap:NUHMregion}

Solving the RECCs, which is derived from the given 1loop RG equations in \cite{Martin:1993zk} as in Appendix \ref{chap:MSSM}, we evaluate the Taylor coefficients of the conditions, Eqs.(\ref{eq:MaC1}), (\ref{eq:MaC2}) and (\ref{eq:MaC3}), in terms of the fundamental parameters \Eq{fp}. Directly solving these constraints, we obtain $\tl{\mathcal{M}}_f^{\rm fund}$, namely the whole viable region of the NUHM.

\begin{figure}[h]
\begin{center}  
\begin{minipage}{0.49\hsize}
   \includegraphics[width=50mm]{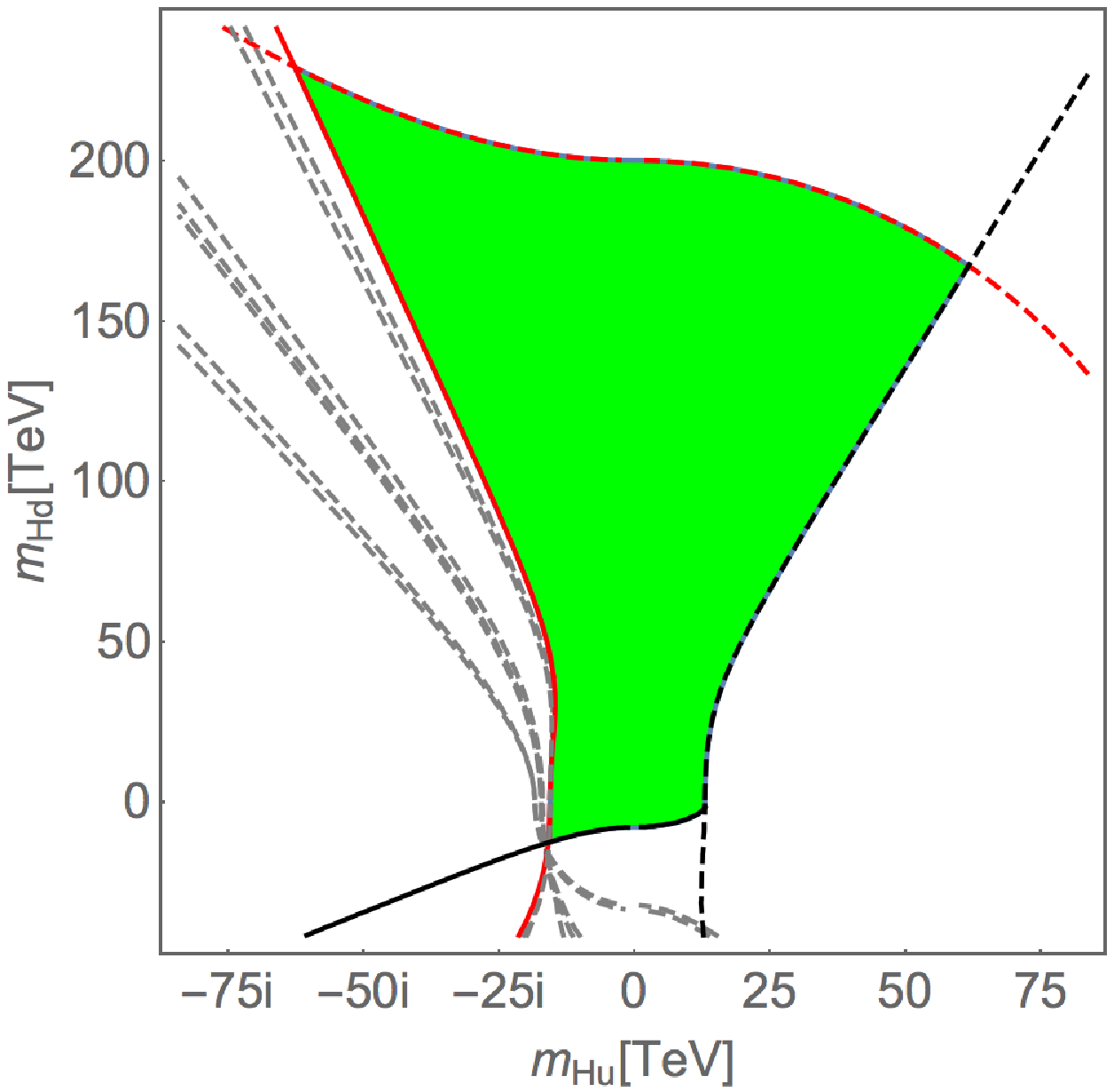}
\end{minipage}
\begin{minipage}{0.49\hsize}
  \includegraphics[width=77mm]{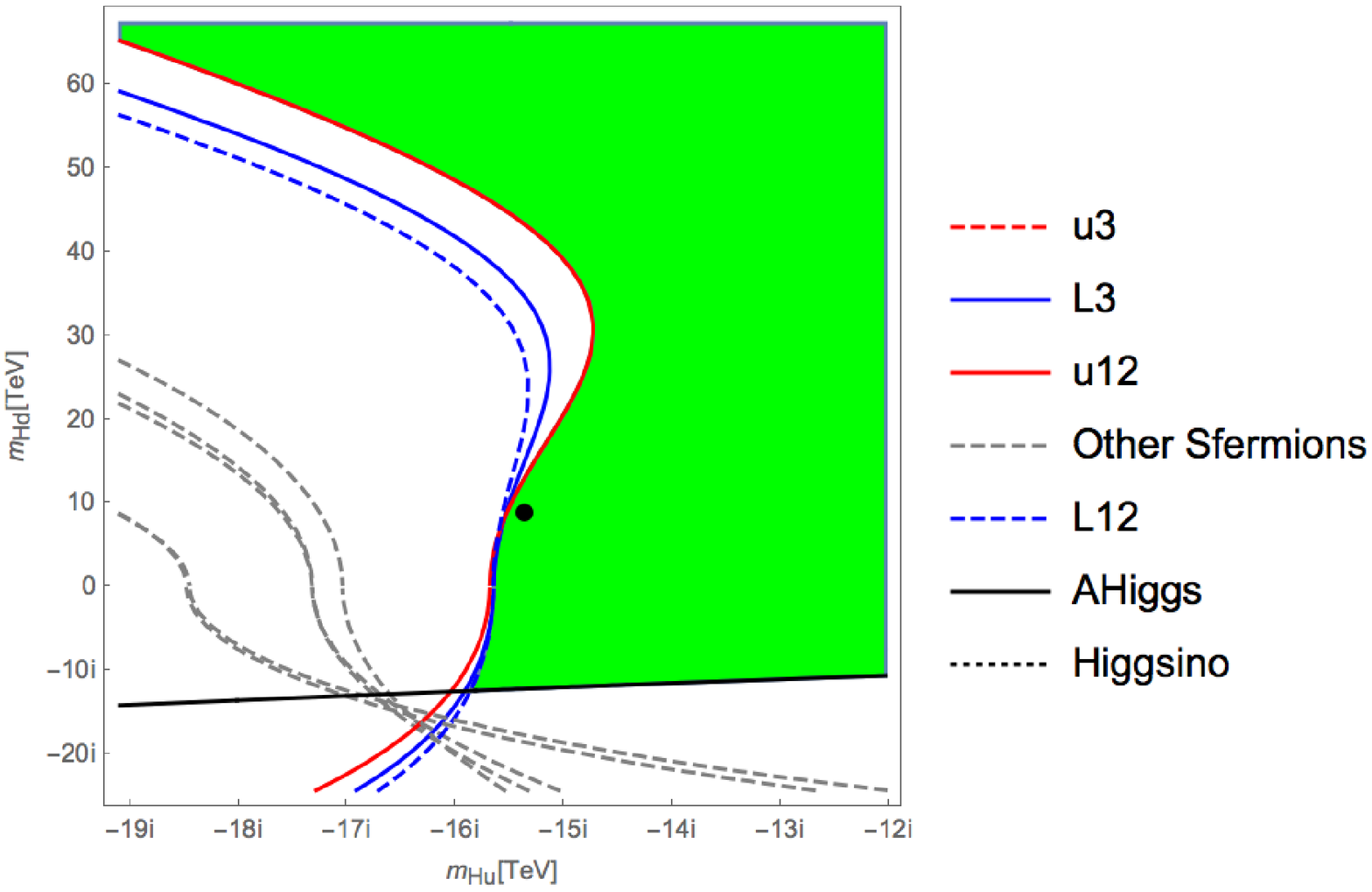}
  \end{minipage}
      \end{center}
\caption{The viable region and the boundaries of the experimental constraint on the slice of $M_0$=750GeV, $A_0$=500GeV with $\tan\beta$=10 (left). The region with green is experimentally allowed with the measured Higgs boson mass and a correct electroweak vacuum. Magnified view near an outline is presented in the right-hand side, where the boundary profiles are shown in detail.   
The black dot, ($-$15356$i$GeV, 8836GeV), represents the low energy parameters given in Table.\ref{tab:ILSQ}}
 \label{fig:wholerg}
 \end{figure}

$\tl{\mathcal{M}}_f^{\rm fund}$ is characterized by four independent parameters, as the seven fundamental parameters $G_a$ are constrained by these three equations in Eqs.(\ref{eq:MaC1}) and (\ref{eq:MaC2}). A two dimensional slice in the four dimensional viable region, $\mathcal{M}^{\rm fund}_f$, is shown in Fig.\ref{fig:wholerg}. 
Also shown are the mass bounds of superparticles and the pseudoscalar Higgs boson (A Higgs), where we have used indices ``12" and ``3" to denote the 1st/2nd and 3rd generations, respectively.

There is a new interesting region near the red solid line in Fig.\ref{fig:wholerg}, the boundary of which 
is given by the mass bound of the first two generation up-type squarks. 
We note that the stops are kept heavy to reproduce the measured Higgs boson mass.
We call this region an inverted light squark (ILSQ) region. Here ``inverted" stands for the new superparticle 
mass pattern characterizing this region where the first two generation squarks are light, contrary to the ordinary 
light stop. 
Table.\ref{tab:ILSQ} illustrates the low energy parameters corresponding to the point represented by the black 
dot shown in Fig.\ref{fig:wholerg}, where the up-type squarks in the first two generations may be within the reach of forth coming experiments. 
We have confirmed that the Higgs boson mass 124(1)GeV is evaluated from FeynHiggs 2.10.2 
\cite{Hahn:2013ria} with the parameters in Table.\ref{tab:ILSQ} as an input. We have also confirmed that the 
ILSQ region still exists including 2loop RG running by solving RECCs at the two-loop level.

Furthermore, we can observe a special point in the left-hand side of Fig.\ref{fig:wholerg}, where most of the contours including the gray dotted lines concentrate. Since they correspond to the mass bounds for the sfermions, the concentrating point, if near enough to the viable region, implies a surprising possibility. That is most of the scalar masses are just above the experimental bounds in spite of the heavy stops which weigh around 6TeV. 
However, as in the right figure the approximately concentrating point is too far away from the viable region, e.g. $\O((10\TEV)^2)$ in mass squareds.
In fact, the point is not excluded by the experimental mass bounds but excluded due to the unstable electroweak vacuum namely the A Higgs is tachyonic, $m_A^2<0$. 
We will discuss the instability in detail in Sec.\ref{chap:ILSQ}, and show the instability can be alleviated with large $\tan \beta$ in Sec.\ref{chap:g21sigma}.

\begin{table}[!t]
\caption{The low energy parameters corresponding to the black dot,
($m_{\rm Hu0}$, $m_{\rm Hd0}$, $M_0$, 
$A_0$)=($-$15400$i$GeV, 8840GeV, 750GeV, 500GeV) with $\tan\b=10$, shown in Fig.\ref{fig:wholerg}. The other fundamental parameters, ($m_0$, $\mu_0$, $B_0$)=(3200GeV, 12600GeV, 1800GeV), are evaluated from the solution of Eqs.(\ref{eq:MaC1}) and (\ref{eq:MaC2}).  ``EW" stands for electroweak.
} 
\label{tab:ILSQ}
\begin{center} \begin{tabular}{|c|c|c|c|c|c|c|c|c|c|c|c|c|c|}
\hline
EW scale&$m_{{\rm H_u}}$&$m_{{\rm H_d}}$&$m_{{\rm \tilde{Q}}3}$&$m_{{\rm \tilde{u}}3}$&$m_{{\rm \tilde{d}}3}$&$m_{{\rm \tilde{L}}3}$&$m_{{\rm \tilde{e}}3}$&$m_{{\rm \tilde{Q}12}}$\\ \hline
TeV&  $-$12.1$i$&8.3&6.6&7.5&4.3&1.3&5.1&4.1\\ \hline
$m_{{\rm \tilde{u}12}}$&$m_{{\rm \tilde{d}12}}$&$m_{{\rm \tilde{L}12}}$&$m_{{\rm \tilde{e}12}}$&$M_1$&$M_2$&$M_3$&$\mu$&$m_A$\\ \hline
1.6&4.4&1.4&5.2&0.31&0.62&2.2&12.2&14.9\\ \hline 
\end{tabular} \end{center}
\end{table}

\subsection{Mechanism for the inverted light squark}

\label{chap:ILSQ}

In this subsection we will explain how the characteristic mass pattern in the ILSQ region is generated in spite of the universal sfermion mass condition, \Eq{unis}, and the requirement of heavy stops, \Eq{MaC2}. Since these two conditions are imposed at two different scales, the RG running should be essential.
The RG equation for a right-handed up-type squark mass is,
\begin{align}
\label{eq:RGEu}
{d \over dt}{m_{\rm \tilde{u}i}^2}& \sim \frac{2}{16 \pi^2} \biggl \lbrace 2 y_{\rm t}^2X_{\rm t}\delta_{ i3}+Yg'^2 S-{16 \over 3}g_3^2M_3^2-{4}g'^2Y^2M_1^2\biggr \rbrace,\\
\laq{S}
S \equiv &\left( m_{{\rm {\rm H_u}}}^2-m_{{\rm H_{{\rm d}}}}^2+ {\rm Tr}[m_{{\rm \tilde{Q}}}^2 -m_{{\rm \tilde{L}}}^2-2m_{{\rm \tilde{u}}}^2+ m_{{\rm \tilde{d}}}^2 +m_{{\rm \tilde{e}}}^2] \right),\\
\label{eq:Yukawa}
X_{\rm t} \equiv & m^2_{\rm H_u}+m_{\rm \tilde{Q}_3}^2+m_{\rm \tilde{u}_3}^2+\ab{A_{\rm t}}^2,
\end{align}
where $i$ represents 12 or 3, and $Y$ is the hypercharge, $-$2/3 \cite{Martin:1993zk}. $y_t$ is the top-
Yukawa 
coupling while $g'$ and $g_3$ are gauge couplings of $\rm U(1)_Y$ and $\rm SU(3)_c$, respectively. The 
split between the 3rd and 1st/2nd 
generation up-type squarks should originate from the Yukawa term, the first term in \Eq{RGEu}. To raise the stop mass $m_{\tl{u}
3}$ large enough for the inverted hierarchy, we need a 
negative and large $X_t$ that dominates over the other terms in \Eq{RGEu}. This is realized when $m_{\rm 
H_u}^2$ is large and negative.

For sufficiently large $\tan\beta$, the bottom and tau Yukawa couplings can be effectively large in spite of the observed small mass ratio of bottom/tau to top. If $m_{\rm Hd}^2$ is large and negative, the same 
argument applies to down-type squarks and sleptons. 
In summary, the inverted hierarchy can be generated by the Yukawa contribution through the RG running due to a negative and large Higgs mass squared parameter.

Furthermore, we would like to explain the sfermion mass splitting within a generation. This originates in the gauge interactions, especially for $\rm U(1)_Y$ gauge symmetry. In particular, the value of $S$ in \Eq{S} can largely deviate from zero in the NUHM contrary to the CMSSM case where $S$ is always zero. We find that when our mechanism of the ILSQ works, $S$ is non-zero in order to avoid the unstable electroweak vacuum, unless $\tan\beta$ is substantially large. We show this by reductio ad absurdum.  
Suppose that $S=0$ is satisfied with a large and negative $m_{\rm Hu0}^2$, namely $m_{\rm Hu0}^2=m^2_{\rm Hd0}\ll 0$ at the GUT scale.
Since $X_t$ dominates over the other terms in \Eq{RGEu}, 
the RG equations of $m^2_{\rm H_u}$ and $m^2_{\rm H_d}$ are also controlled by the large and negative Higgs masses \cite{Martin:1993zk}, 
\beq
\laq{betamHu2}
\b_{m_{\rm Hu}^2} \sim {1 \o 16\pi^2 } 6y_t^2 X_t^2 \sim {1 \o 16\pi^2 } 6y_t^2 m_{\rm Hu}^2,
\ebq
\laq{betamHd2}
\b_{m_{\rm Hd}^2} \sim {1 \o 16\pi^2 } \( 6y_b^2 m_{\rm Hd}^2+2y_\t^2  m_{\rm Hd}^2 \).
\eeq
Hence, the RG running effect decreases the absolute value of $m_{\rm Hu}^2$ at the experimental scale, while $m_{\rm Hd}^2$ does not change so much due to the smaller Yukawa couplings. 
This implies that at low energy 
\eqn{ \laq{tinst}0\gg m_{\rm Hu}^2 > m_{\rm Hd}^2}
is satisfied and the problem of tachyonic instability of the electroweak vacuum occurs, as $m_A^2<0$ by solving Eqs.\eq{MaC1}. Therefore, we need a greater $m_{\rm H{d0}}^2$ to stabilize the electroweak vacuum and hence $S<0$. 
Notice that this argument depends on the sign of the Higgs mass squareds and the size of $\tan\b$.
Since \Eq{betamHu2} decreases the absolute value of $m_{\rm Hu}^2$, if $S$ is cancelled by positive and large $m_{\rm Hu0}^2$ and $m_{\rm Hd0}^2$, \Eq{tinst} is replaced by $0 \ll m_{\rm Hu}^2 < m_{\rm Hd}^2$ and the instability problem does not occur. This is the reason why the CMSSM is allowed to have a stable electroweak vacuum. Also if $\tan \beta$ is large enough, the RG running of $m_{\rm Hd}^2$, \Eq{betamHd2}, can be effective and the vacuum instability problem should be alleviated with $S\sim 0$ in the ILSQ region.

For not so large $\tan\beta$, the property of $S<0$ in the ILSQ region makes $\tl{u}_{12}$ the lightest squark as in Table.\ref{tab:ILSQ}, because it has the least hypercharge.
The requirement of $S<0$ is also the reason why the approximately concentrating point in the right-hand side of Fig.\ref{fig:wholerg} is far away from the viable region. In fact, the approximately concentrating point lies on the line of $S=0$ where the first two generation sfermion 
masses can be small without splitting. 
On the other hand, since $S\sim 0$ may be allowed when $\tan \b$ is sufficiently large, in this case 
there is a possibility to have an ILSQ region including most of the first two sfermion masses just above the experimental bounds. This is the case in \Sec{g21sigma}.\\

As we have discussed, the ILSQ region has a new superparticle mass pattern essentially related to the non-universal Higgs masses. Therefore, this mass pattern should be one of the most characteristic phenomenon of the NUHM, as the CMSSM never realize this.

\section{Terribly Fine-Tuned but Important Region}
\label{chap:muong-2}

In Sec.\ref{chap:NUHM}, we have presented the whole viable region of the NUHM by solving the 
experimental constraints in terms of the fundamental parameters which can be obtained from the method advocated in \Sec{approach}.  
We have found a new phenomenologically interesting region near the boundary corresponding to a squark mass 
bound. However, an interesting region with a characteristic property is not necessarily around the outlines, 
rather inside the viable region. This is particularly the case when the characteristic property requires a fairly 
large amount of fine tuning among the parameters. Notice that it would be difficult for the scatter plot method to 
identify such a fine-tuned region. \\

 In this 
section we will show how to analyze a large and complicated viable region to find out the characteristic 
properties localized inside of it, and argue this analysis has an advantage when applied to a FT-SUSY.

To illustrate the idea, consider here the muon anomalous magnetic dipole moment (muon $g-2$), $\alpha_\mu$. The muon $g-2$ anomaly is a hint of new physics,
as the discrepancy between the theoretical and experimental values exceeds the $3~\sigma$ level of these errors \cite{Bennett:2006fi}, \cite{Hagiwara:2011af}, \cite{Davier}
\begin{eqnarray}
\laq{g2dev}
\alpha_{\rm \mu}^{\rm exp} -\alpha_{\rm \mu}^{\rm SM} = (26.1\pm 8.0)\times 10^{-10}.
\end{eqnarray}
In the generic MSSM at least three light superparticles, weigh around 300GeV, are needed to generate a large enough contribution \cite{Moroi:1995yh}, \cite{Lopez:1993vi}. Therefore, the region of the generic MSSM, $\tl{\mathcal{M}}_e^{\rm gen}$, that fills the discrepancy of the muon $g-2$, should be characterized by three parameters with very tiny values compared to the stop mass parameters around $6\TEV$. Therefore, if exists, the region of interest of the NUHM now, $\tl{\mathcal{M}}_f^{\rm fund}$, should be fine-tuned from the whole viable region in the previous section.

The particular diagram we consider for the SUSY contribution to the muon $g-2$ is made of a loop including a bino and smuons of both chiralities \cite{Moroi:1995yh}, \cite{Lopez:1993vi}. 
Following \cite{Moroi:1995yh}, we obtain an approximated formula of the SUSY contribution to the muon $g-2$,
\begin{align}
\label{eq:g2loop}
\delta \alpha_{\rm \mu} &\equiv \alpha_{\rm \mu}^{\rm MSSM}-\alpha_{\rm \mu}^{\rm SM} \nonumber \\
&\sim \frac{1}{16\pi ^2}\frac{g'^{2} m_{\mu}^2 M_1 \mu \tan \beta \min{\{ m_{\rm \tilde{L}12}^2,m_{\rm \tilde{e}12}^2,M_1^2\}}}{2m_{\rm \tilde{L}12}^2m_{\rm \tilde{e}12}^2M_1^2} \nonumber \\ 
&\sim 5 \times 10^{-10} \frac{(100{\rm GeV})^2 M_1 \mu \tan \beta \min{\{m_{\rm \tilde{L}12}^2,m_{\rm \tilde{e}12}^2,M_1^2\}}}{m_{\rm \tilde{L}12}^2m_{\rm \tilde{e}12}^2M_1^2}.
\end{align} 
This rough approximation will be corrected by fitting the results evaluated in FeynHiggs 2.11.2 \cite{Hahn:2013ria} by varying the overall coefficient of this function.

\subsection{How large can the muon $g-2$ be in the NUHM?}
\label{chap:maxg2}
 In order to clarify whether the NUHM can explain the muon $g-2$ anomaly, an efficient way is to evaluate the maximal value of the muon $g-2$ in the NUHM. This immediately draws a conclusion of whether the muon $g-2$ anomaly can be explained. 
Therefore we impose a condition,
\begin{align}
\label{eq:g2MaC1}
{\rm maximize}{[\delta \alpha_{\rm \mu}]} \ &{\rm by \ varying \ } m_0,
\end{align}
in addition to Eqs.(\ref{eq:MaC1})--(\ref{eq:MaC3}). In fact, we can vary all the free parameters to maximize the muon $g-2$, however, for the illustrative purpose we only vary one parameter. This reduces a free parameter, $m_0$, in $\mathcal{M}_f^{\rm fund}$.

\begin{figure}[!t]
\begin{minipage}{0.49\hsize}
  \begin{center}
   \includegraphics[width=50mm]{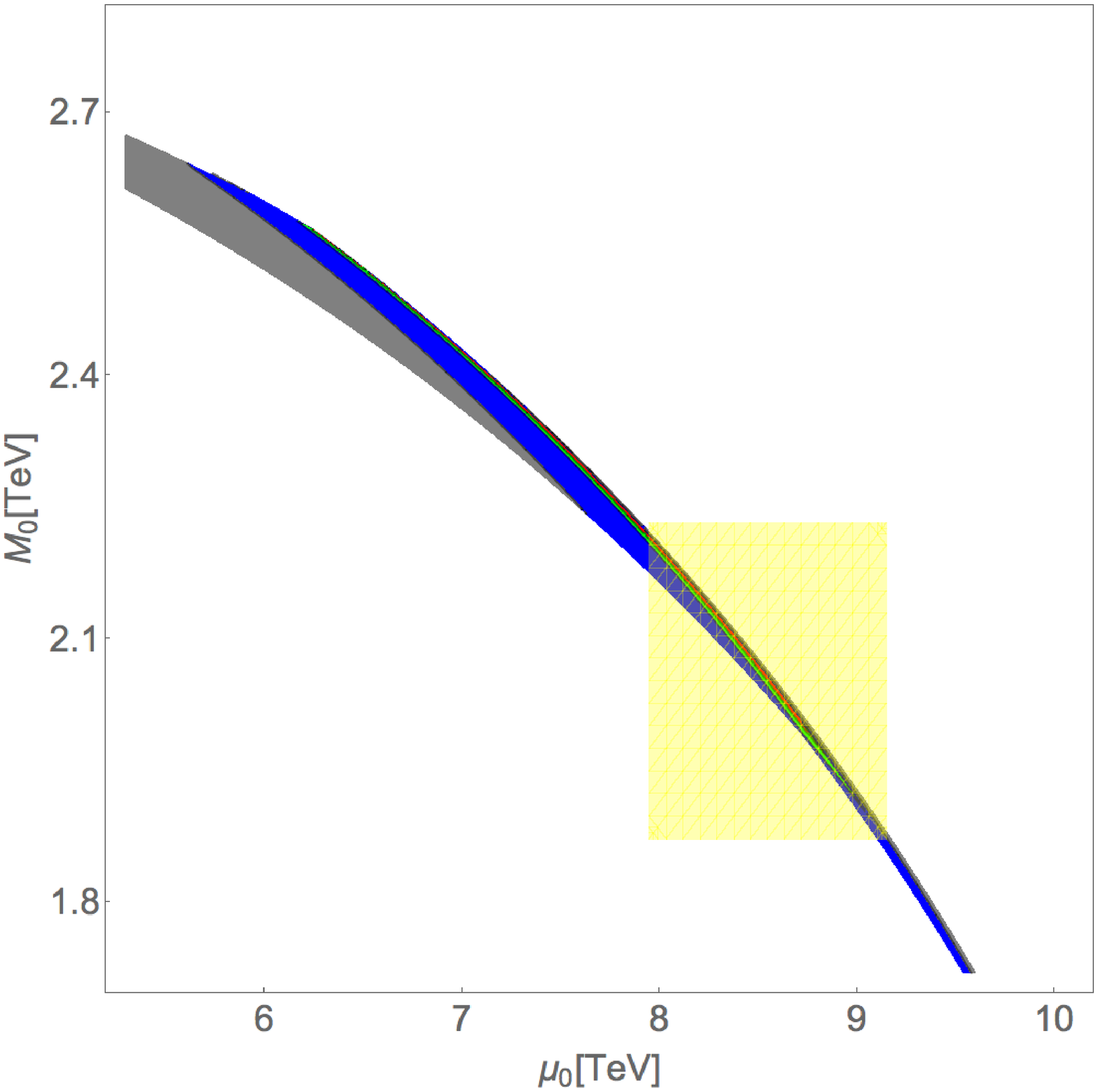}
 \end{center}
 \end{minipage}
\begin{minipage}{0.49\hsize}
  \begin{center}
   \includegraphics[width=50mm]{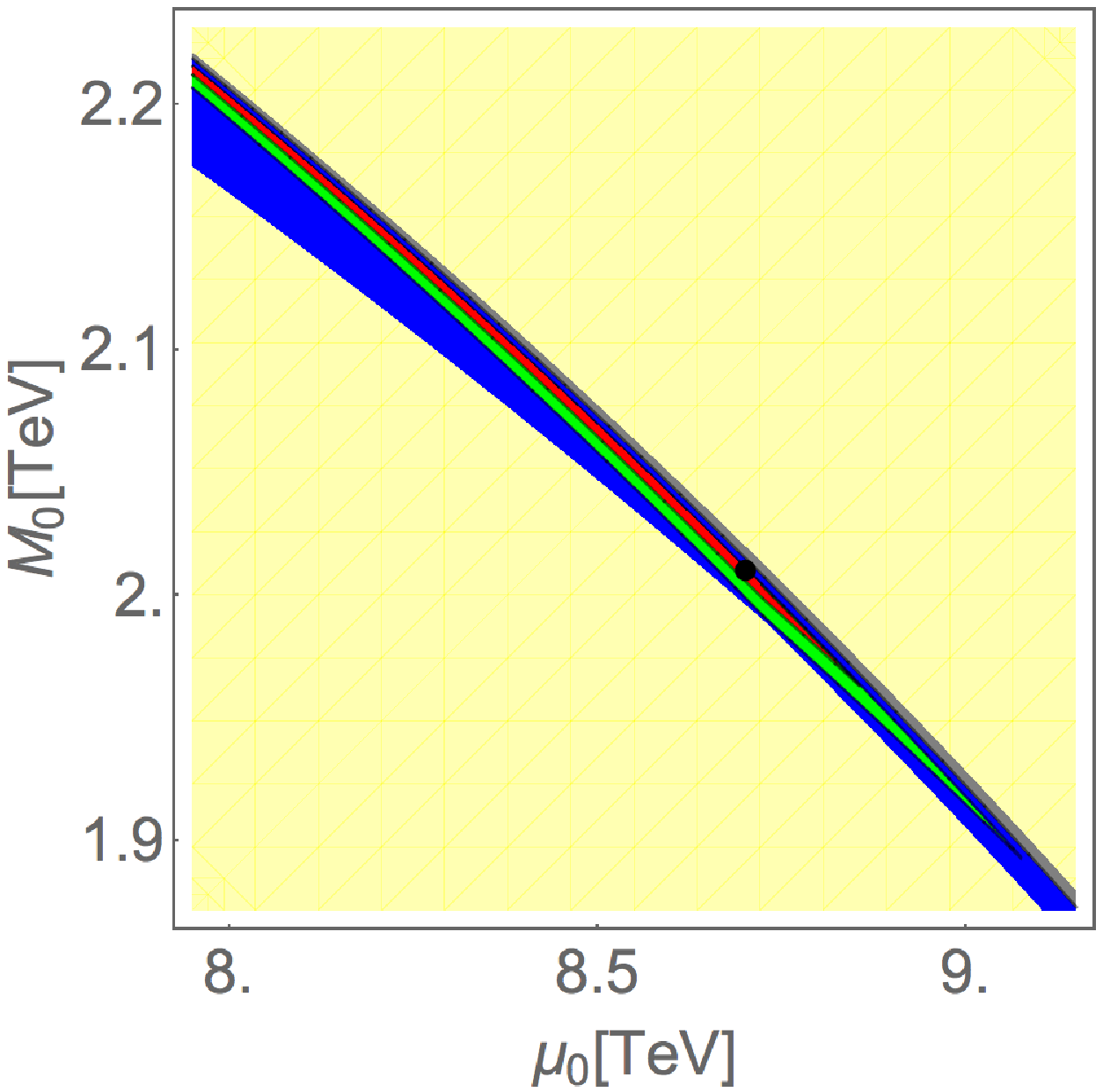}
 \end{center}
 \end{minipage}
 \caption{ The contour plot of the maximized muon $g-2$ by varying $m_0$ on a slice at $A_0=0$ with $\tan\beta$=35 (left). Magnified view of the yellow region is presented in the right-hand side. The red (green, blue) region represents that the total muon $g-2$, including the maximal NUHM correction, is within the $1~\sigma$ ($2~\sigma$, $3~\sigma$) level error of the observed value. The black dot, (8700GeV, 2010GeV), represents the low energy parameters given in Table.\ref{tab:muong235}.}
\label{fig:maxg2}
\end{figure}

By solving the conditions, Eqs.(\ref{eq:MaC1})--(\ref{eq:MaC3}) and (\ref{eq:g2MaC1}), in terms of $A_0, M_{0}$, and $\m_0$, we obtain a three dimensional region of interest, $\mathcal{\tl{M}}_f^{\rm fund}$. An $A_0=0$ slice of the solution is shown in Fig.\ref{fig:maxg2} with $\tan\beta=35$. The contours represent the maximized total muon $g-2$ by varying $m_0$.

We find the maximized muon $g-2$ can exceed $18.1 \times 10^{-10}$ with $\tan\beta \gtrsim 25$, namely the NUHM is able to fill the discrepancy of the muon $g-2$ within the $1~\sigma$ level error.

In fact, there are two kinds of regions, namely type 1 and type 2 regions, depending on the lightest particle in the loop diagram corresponding to \Eq{g2loop}. The type 1 region explains the muon $g-2$ anomaly with either $\tl{e}_{12}$ or $\tl{L}_{12}$ as the lightest particle in this diagram, while the type 2 region has bino as the lightest particle. 
Fig.\ref{fig:maxg2} corresponds to the former one, while the 
region of type 2 appears for $\tan\beta \gtrsim 55$ as we will show in \Sec{g21sigma}.

\begin{table}[!h]
\caption{The low energy parameters corresponding to the black dot, $(M_0,\mu_0,A_0)$=(2010GeV, 8700GeV, 0GeV), in Fig.\ref{fig:maxg2} with $\tan\beta$=35. The other fundamental parameters, $(m_0, m_{\rm Hu0}, m_{\rm Hd0}, B_0)$ = (1186$i$GeV, 9375$i$GeV, 8326$i$GeV, $-$768GeV), are the solution of Eqs.(\ref{eq:MaC1}), (\ref{eq:MaC2}) and (\ref{eq:g2MaC1}).}
\label{tab:muong235}
\begin{center} \begin{tabular}{|c|c|c|c|c|c|c|c|c|}
\hline
EW scale &$m_{{\rm H_u}}$&$m_{{\rm H_d}}$&$m_{{\rm \tilde{Q}}3}$&$m_{{\rm \tilde{u}}3}$&$m_{{\rm \tilde{d}}3}$&$m_{{\rm \tilde{L}}3}$&$m_{{\rm \tilde{e}}3}$&$m_{{\rm \tilde{Q}12}}$\\ \hline
GeV&8044$i$&7992$i$&5834&6161&5089&1637&2275&5168\\ \hline \hline
$m_{{\rm \tilde{u}12}}$&$m_{{\rm \tilde{d}12}}$&$m_{{\rm \tilde{L}12}}$&$m_{{\rm \tilde{e}12}}$&$M_1$&$M_2$&$M_3$&$A_{{\rm u3}}$&$A_{{\rm d3}}$\\ \hline
4920&4999&429&429&839&1659&5770&$-$4497&$-$6749\\ \hline \hline
$A_{{\rm e3}}$&$A_{{\rm u12}}$&$A_{{\rm d12}}$&$A_{{\rm e12}}$&$\mu$&$m_{\rm A}$ &FeynHiggs& $m_{\rm h}$ by FH& $\delta \alpha_\mu$ by FH \\ \hline 
$-$1011&$-$7891&$-$7820&$-$1372&8043&907 & (2.11.2) \cite{Hahn:2013ria}&126(1.4)GeV&$2.5\times 10^{-9}$\\ \hline \end{tabular} 
\end{center}

\end{table}

\subsection{The region explaining the muon $g-2$ anomaly and the mechanism}

\label{chap:g21sigma}
Since the NUHM has a region where the muon $g-2$ anomaly is explained, now we would like to explore the features of this region.
Since we know that an outline may have some phenomenological information about the nearby region,
it is meaningful to show the outlines of the region where the muon $g-2$ is just at the experimental central value, rather than the maximal value.
We impose a condition instead of \Eq{g2MaC1},
\begin{align}
\label{eq:g2MaC1sigma}
\frac{(100{\rm GeV})^2 M_1 \mu \tan \beta \min{\{ m_{\rm \tilde{L}12}^2,m_{\rm \tilde{e}12}^2,M_1^2 \}}}{m_{\rm \tilde{L}12 }^2m_{\rm \tilde{e}12}^2M_1^2} =\left\{
\begin{array}{ll}
 &10 \ (\tan \beta \sim35)\\
&7.5 \ (\tan \beta \sim 60).
\end{array} \right.
\end{align}
The factor 10 and 7.5 in the right-hand side are fitted experimentally using FeynHiggs 2.11.2 \cite{Hahn:2013ria}.

\begin{figure}[!t]
\begin{center}
 \begin{minipage}{0.45\hsize}
   \includegraphics[width=70mm]{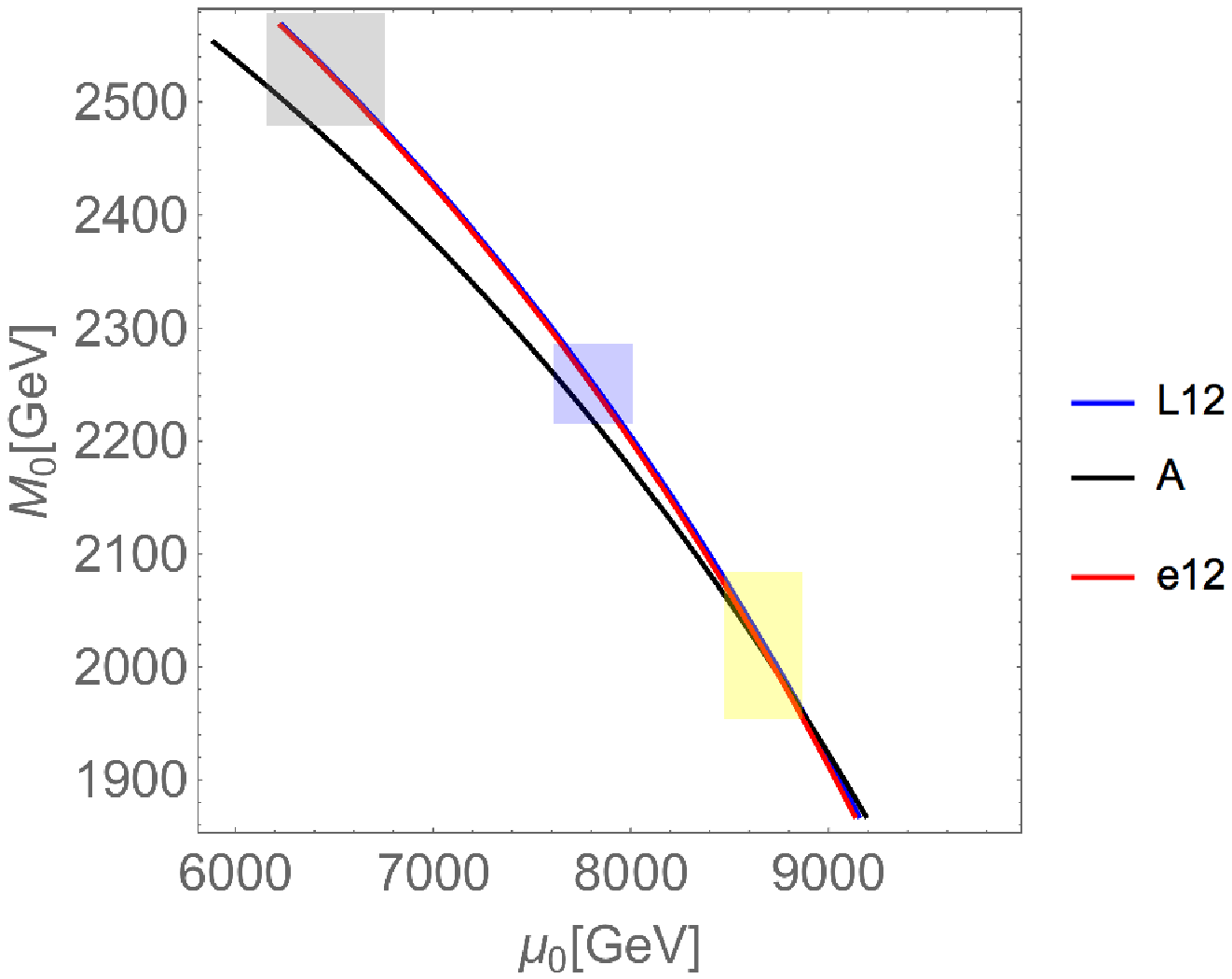}
\end{minipage}
 \begin{minipage}{0.45\hsize}
   \includegraphics[width=70mm]{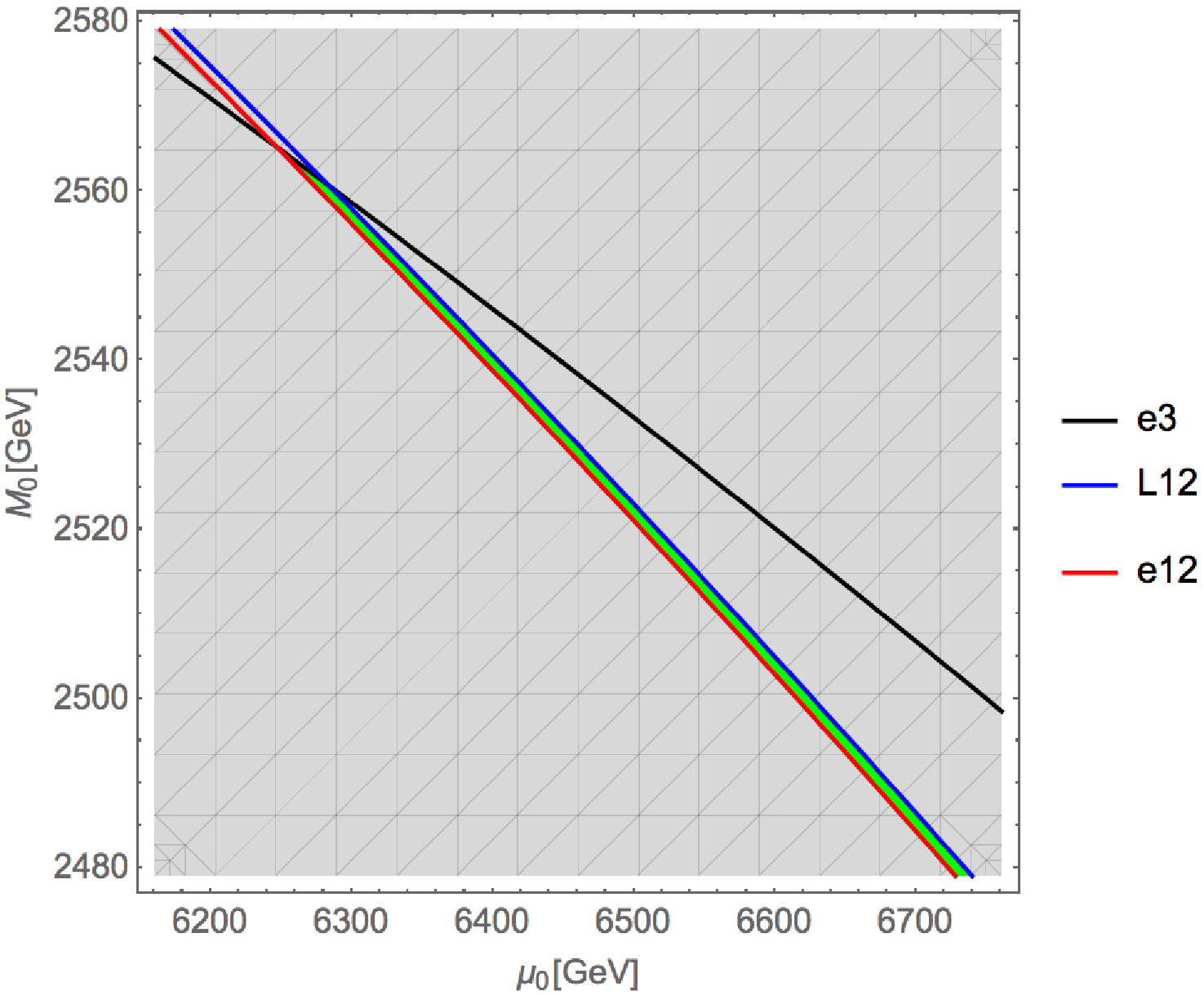}
 \end{minipage}
 \end{center}
\begin{center}
 \begin{minipage}{0.45\hsize}
   \includegraphics[width=70mm]{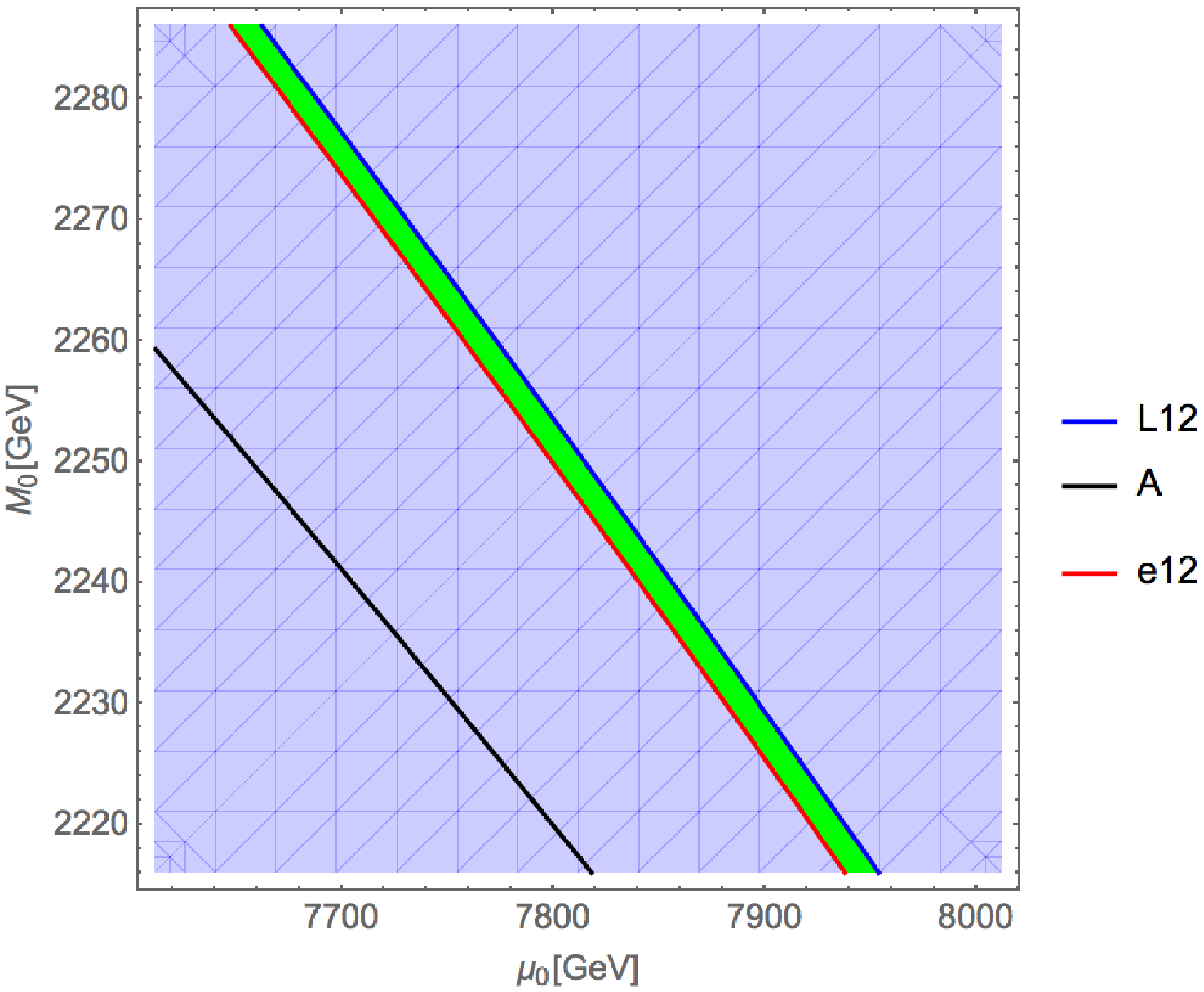}
 \end{minipage}
 \begin{minipage}{0.45\hsize}
   \includegraphics[width=70mm]{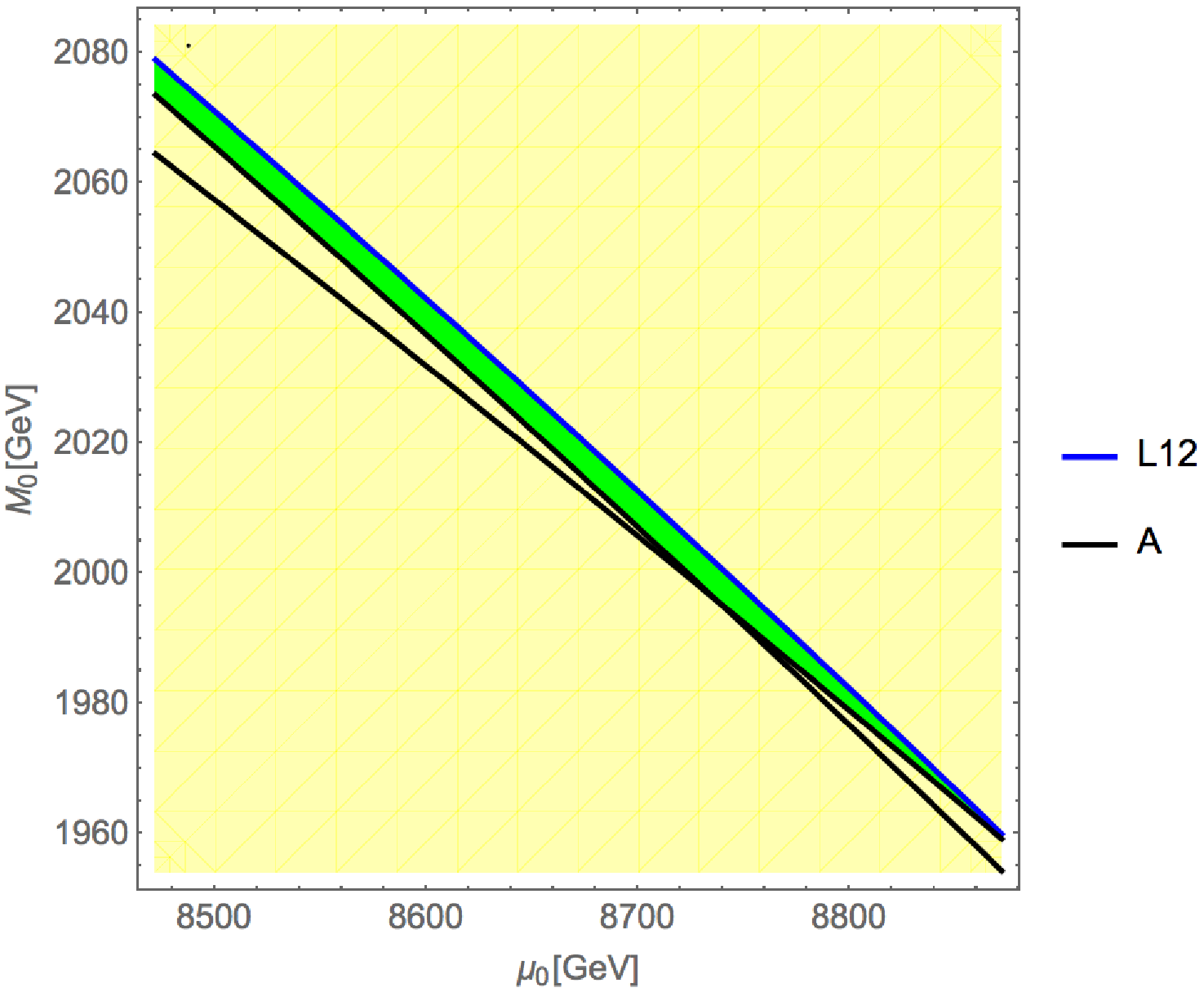}
 \end{minipage}
  \end{center}
\caption{$A_0$=0 slice of the region of type 1 with $\tan\beta$=35. The region with green 
explains the muon $g-2$ anomaly within its $1\s$ level error where one of the smuons 
is lighter than the bino.
 The magnified view of the corresponding colored region is also shown with the boundary profiles in detail. 
 The profiles of the boundaries that do not constitute the outlines are turned off. }
 \label{fig:g21sigma35}
\end{figure}

\begin{figure}[!t]
\begin{center}
   \includegraphics[width=100mm]{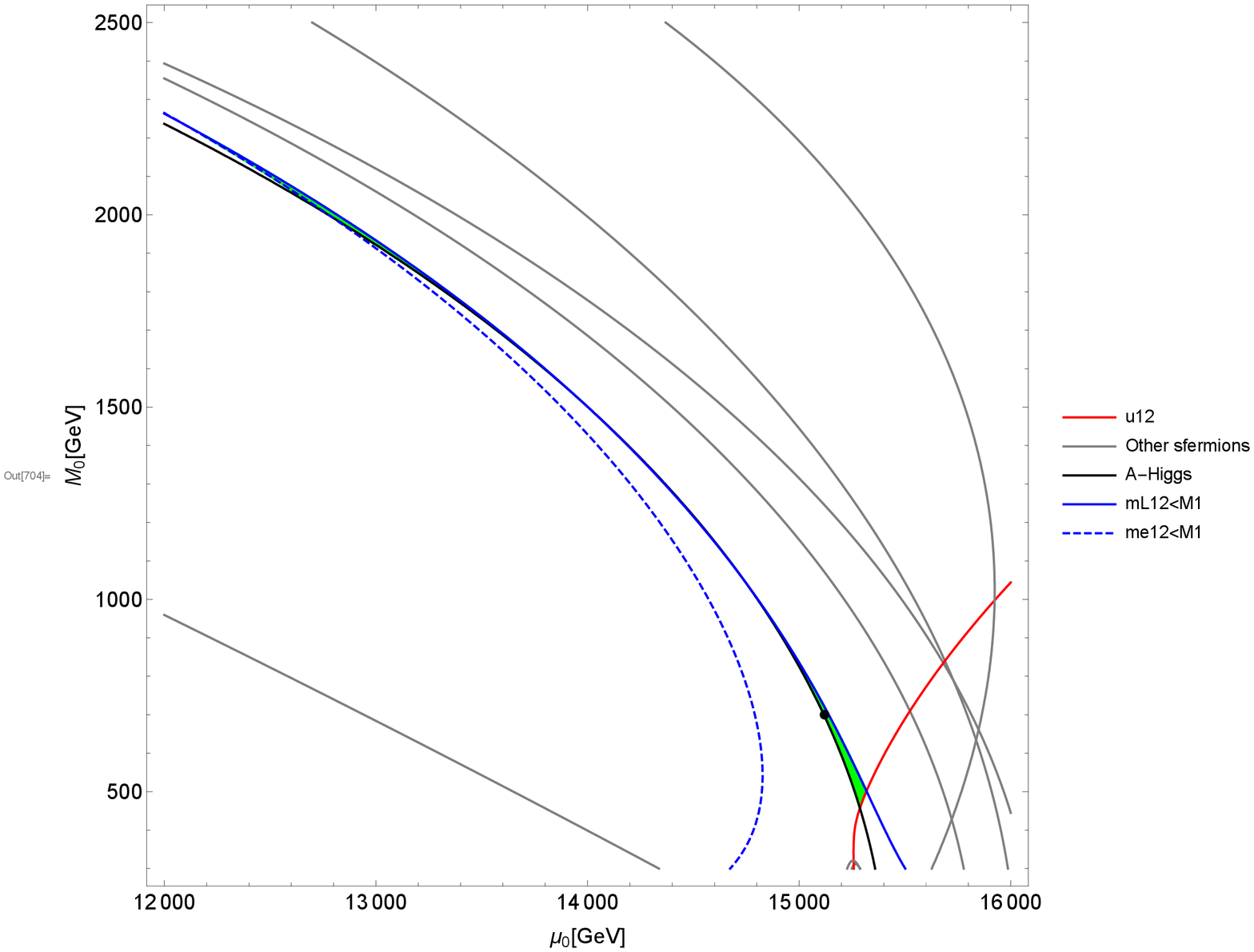}
  \end{center}
 \caption{$A_0$=0 slice of the region of type 2 with $\tan \b =60$. The region with green explains the muon $g-2$ anomaly within its $1\s$ level error where the bino is lighter than the smuons.
  The boundary profiles are also shown. The blue solid and dashed lines denote the boundary of the condition to have type 2 region which constrains the bino to be the lighter than the smuons.
 The black dot, (15120GeV, 700GeV), has low energy parameters given in Table.\ref{tab:g2bino}}
\label{fig:g2bino}
\end{figure}

The solutions by solving Eqs.(\ref{eq:MaC1})--(\ref{eq:MaC3}), and (\ref{eq:g2MaC1sigma}) in terms of $A_0, M_{0}$, and $\m_0$, are presented
 in Fig.\ref{fig:g21sigma35} and Fig.\ref{fig:g2bino} corresponding to type 1 and type 2 regions, respectively. Also shown are the boundary profiles.

If the muon $g-2$ anomaly is explained by the region in Fig.\ref{fig:g21sigma35}, from the boundary profiles 
some of light smuons, selectrons, right-handed stau and A-Higgs might be within the reach of the forth coming 
experiments. On the other hand, Fig.\ref{fig:g2bino} implies that many superparticles might be within the reach 
of the forth coming experiments, as many boundaries of the experimental constraints are within the distances of $\lesssim\O(\TEV)$ from this interesting region.  A set of low 
energy parameters corresponding to the type 2 region is given in Table.\ref{tab:g2bino}. We have confirmed that the Higgs mass and the muon $g-2$ anomaly are explained using FeynHiggs 2.10.2 \cite{Hahn:2013ria} 
with an input of these parameters. We have also confirmed that the parameter region does not vanish but shifts to a different area in the parameter space by solving the RECCs at the 2loop level with some of the 1loop threshold 
corrections included in the conditions Eqs.\eq{MaC3}. Chosen in this parameter region, some sets of fundamental parameters are inputted into SOFTSUSY 3.7.1 \cite{Allanach:2001kg} and a similar pattern of the low energy 
parameters to those in Table.\ref{tab:g2bino} is produced. For example, with an input $\{\tan\b=61, M_{\rm 0}=919 \GEV, m_{\rm Hd0}^2=-1.835\times 10^{8} \GEV^2, m_{\rm Hu0}^2=-2.200\times 10^{8}\GEV^2, m_0=623.0 \GEV, \mu_0>0 \}$, SOFTSUSY 3.7.1\footnote{We have set the number of loops in the Higgs boson mass computation to be one, as otherwise the computation would not converge due to the terrible fine tuning.} shows a set of low energy parameters
 from which the measured Higgs boson mass and the muon $g-2$ within the 1$\s$ error are evaluated by FeynHiggs \cite{Allanach:2001kg,Hahn:2013ria}.\\

Now we would like to consider how the muon $g-2$ anomaly is explained. From \Eq{g2MaC1sigma}, we find that large $\mu-$term and $\tan\b$ are favored for a large value of
the muon $g-2$. This requires large and negative Higgs masses to have a correct electroweak vacuum 
due to Eqs.\eq{MaC1}. Therefore the mechanism of the ILSQ in \Sec{ILSQ} can be applied. As noted, a negative $S-$term is required to stabilize the electroweak vacuum when $\tan\beta$ is 
not substantially large. Neglecting RG running due to gauginos, $\tl{u}_{12}$ becomes the lightest sfermion 
in most of the cases as in Table.\ref{tab:ILSQ}. This excludes a smuon around 300GeV, as it should be 
heavier than $\tl{u}_{12}$ which is bounded below by 1TeV. If we would like to decrease the smuon masses in the 
ILSQ region, what we should do is to increase the gluino mass.
The RG effect from a heavy gluino, as the third term in \Eq{RGEu}, raises the masses of squarks universally from those of sleptons.  However the universal gaugino mass condition, \Eq{unig}, implies a heavy bino. Hence, 
the lightest particle in the loop diagram corresponding to \Eq{g2loop} becomes a smuon which can be seen from the boundary profiles in Fig.
\ref{fig:g21sigma35}. This is the case for type 1 region.

The type 2 region where bino is the lightest particle in the loop diagram corresponding to \Eq{g2loop} can be realized with sufficiently large $\tan\b$. This is consistent with the argument noted in \Sec{ILSQ}: $S\sim 0$ can be realized with large $\tan\beta$. In this case, we do not need a heavy gluino to raise the squark masses and the bino can be as light as the smuons.
Furthermore, all the 1st and 2nd generation sfermions will be light because small gaugino masses and $S-$ term imply the degeneration of the sfermions in the first two generations, and the mechanism of ILSQ works for all of them. This is actually the case in Table.\ref{tab:g2bino}.\\

As we can see in Fig.\ref{fig:maxg2}, Fig.\ref{fig:g21sigma35}, and Fig.\ref{fig:g2bino}, the relevant regions have small sizes due to fine tuning. In fact, fine tuning is already alleviated in these figures as we have reduced a free parameter that is fine-tuned contrary to the stop mass scale by solving \Eq{g2MaC1} or \Eq{g2MaC1sigma}. 
On the other hand, after the discovery of the Higgs boson the muon $g-2$ correction in the NUHM is discussed in several studies \cite{Buchmueller:2014yva} by using the scatter plot method. In these studies, the tiny regions we found in this paper had been overlooked, as the ``pointillism" is not practical to find such a tiny and terribly fine-tuned region. 
Therefore, we have shown that the method advocated in this paper has a strong advantage in identifying a fine-tuned region and this should be an efficient approach to a FT-SUSY.
 
\begin{table}[!t]
\caption{The low energy parameters corresponding to the black dot,
($M_0, \mu_0, A_0$) =(700GeV, 15120GeV, 0GeV), represented in Fig.\ref{fig:g2bino} with $\tan\beta$=60. The other fundamental parameters, ($m_0,m_{\rm Hu0},m_{\rm Hd0},B_0$) =(898GeV, 14910$i$GeV, 13650$i$GeV,21GeV), are obtained by solving the conditions corresponding to Eqs.(\ref{eq:MaC1}), (\ref{eq:MaC2}) and (\ref{eq:g2MaC1sigma}).}
 \label{tab:g2bino}
\begin{center} 
\begin{tabular}{|c|c|c|c|c|c|c|c|c|}
\hline
EW scale &$m_{{\rm H_u}}$&$m_{{\rm H_d}}$&$m_{{\rm \tilde{Q}}3}$&$m_{{\rm \tilde{u}}3}$&$m_{{\rm \tilde{d}}3}$&$m_{{\rm \tilde{L}}3}$&$m_{{\rm \tilde{e}}3}$&$m_{{\rm \tilde{Q}12}}$ \\ \hline 
GeV&11820$i$&11780$i$&6399&7553&4810&4540&6618&2125 \\ \hline \hline 
$m_{{\rm \tilde{u}12}}$&$m_{{\rm \tilde{d}12}}$&$m_{{\rm \tilde{L}12}}$&$m_{{\rm \tilde{e}12}}$&$M_1$&$M_2$&$M_3$&$A_{{\rm u3}}$&$A_{{\rm d3}}$ \\ \hline 
1650&2146&326&1669&292&578&2010&$-$1511&$-$1998 \\ \hline \hline 
$A_{{\rm e3}}$&$A_{{\rm u12}}$&$A_{{\rm d12}}$&$A_{{\rm e12}}$&$\mu$&$m_{\rm A}$&FeynHiggs&$m_{\rm h}$ by FH&$\delta \alpha_\mu$  by FH\\ \hline 
$-$118&$-$2748&$-$2724&$-$478&11820&946&(2.11.2)\cite{Hahn:2013ria}& 124.6(1.3)GeV&$2.1\times10^{-9}$ \\ \hline
\end{tabular} 
\end{center}
\end{table}
\clearpage

\section{Conclusions}

In this paper, we proposed a novel and efficient approach to the
supersymmetric models with some amount of fine tuning, in which the
commonly used approach of scatter plot is inefficient and sometimes
even fails to find relevant regions in the parameter space of
superparticle masses with the limited number of plotted points. The
essential idea of our approach is to directly map the (experimental or
other) constraints at low energy to those in the fundamental parameter
space. We can identify the relevant region in the fundamental
parameter space by filling the interior of the mapped constraints as
the boundaries of the constraints will form the outlines of the
relevant region. Furthermore the areas near the boundaries of the
experimental constraints which are rather easily identified in our
method can be phenomenologically interesting as they will be within
forth coming experiments.

We applied this method to the non-universal Higgs masses (NUHM) model.
The features of the NUHM model are the same as the CMSSM except that
the SUSY-breaking Higgs masses differ from the universal sfermion mass
at the GUT scale. By using our method, we identified the
phenomenologically viable regions of the parameter space and argued
some interesting features of the model. Among other things, we found,
in some cases, that the inverted squark masses are realized, where the
renormalization group effects raise the third generation squark masses
compared to those of the first two generations. This mass pattern is a
characteristic phenomenon of the NUHM model and is never realized in
the CMSSM.

Another application of our method is to identify, within the NUHM
model, the existing but tiny region in the parameter space, where the
SUSY contribution explains the discrepancy of the muon $g-2$ within
the $1~\s$ level of experimental and theoretical errors. The price
to pay is the terrible fine tuning among the parameters, and therefore
the previous studies with the conventional scatter plot method failed to
find this region, drawing misleading conclusions.  This example
illustrates the power of our method in particular when the required
fine tuning is severe. The relevance of our approach will even
increase when the forth coming experiments will give null results in
superparticle searches and more fine tuning will be required to correctly
produce the electroweak scale.

The method advocated in this paper has a variety of applications, some
of which was given in \cite{Shimizu:2015ara} and also will be discussed
elsewhere.

\section*{Acknowledgement}
We would like to thank Yutaro Shoji for collaboration at an early stage of this work. 
This work is supported by the Grant-in-Aid for Scientific Research from the Ministry of Education, Science, Sports, and Culture (MEXT), Japan, No.23104008.

\clearpage

\appendix
\section{RECCs for the generic MSSM}
\label{chap:MSSM}

We would like to show the derivation of RECCs in the generic MSSM with given dimensionless couplings at 
$t_e$. In fact, the following argument can apply to the derivation of RECCs at any loop order with given RG 
equation at the same order. 

The parameters $g_i^f$ and constants of the generic MSSM are classified by their dimensions:
\begin{description}
\item[D=2]Sfermion masses, Higgs masses: $m^2_{i}$
\item[D=1]Gaugino masses, $\mu-$ and $A-$ terms: $M_{i}$
\item[D=0]Gauge couplings and Yukawa couplings: $y_{i}$
\end{description}
where the set of $m^2_i$ is understood to include the off-diagonal elements of usual three by three sfermion mass matrices etc.

A perturbative RG equation for the parameters of dimension $d$ is written in terms of the parameters with dimension$\leq d$:
\begin{align}
\label{RGE2}
\begin{split}
{d \over dt}{m^2_{i}} &=a_{2,i}^{k}(y_{j} )m^2_{k}+a_{1,i}^{kl}(y_{j})M_{k}M_{m},\\
{d \over dt}{M}_{i} &=b_{i}^{k}(y_{j})M_{k}, \\
{d \over dt}{y}_{i} &=c_{i}(y_{j}).
\end{split}
\end{align}
$a^k_{2,i}(y_j),a^{kl}_{1,i}(y_j), b_i^k(y_j)$ and $c_i(y_j)$ are given functions that can be derived from loop calculations in the generic MSSM \cite{Martin:1993zk}. The summation is understood.

Firstly, we can solve the RG equation for dimensionless constants numerically with given $y_{i}$ at $t_e$. 
Substituting this numerical solution for $y_i$, the (effective) RG equation for dimensionful parameters become, 
\begin{align}
\label{eq:effRGE}
\begin{split}
{d \over dt}{m^2_i} &=a_{2,i}^{k}(t)m^2_{k}+a_{1,i}^{kl}(t)M_{k}M_{l}\\
{d \over dt}{M}_{i} &=b_{i}^{k}(t)M_{k}.
\end{split}
\end{align}

Secondly, using \Eq{effRGE}, we can derive the RECCs for condition functions, 
 \eqn{\laq{appcon} \f_2(m_i^2)=m_{1}^2(t_e)-(10^3\GEV)^2~ {\rm and} ~ \f_1(M_i)=M_{1}(t_e)-(10^3\GEV),}
corresponding to simplified experimental mass bounds of a scalar and a fermion, respectively.
Dimensional analysis allows us to guess the forms of the mapped condition functions at an arbitrary scale, $t$, to be
\begin{align}
\label{eq:RECCMSSMd2}
\Phi_2(t) &\equiv \phi_2^{i}(t)m^2_{i}+\phi_2^{ij}(t)M_{i}M_{j}+c_2(t)\\
\label{eq:RECCMSSMd1}
\Phi_1(t) &\equiv \phi_1^{i}(t)M_{i}+c_1(t).
\end{align}
 Therefore, employing Eq.(\ref{eq:gfRECC}) the RECCs for Eqs.(\ref{eq:RECCMSSMd2}) and (\ref{eq:RECCMSSMd1}) are derived as
\begin{align}
\non
{d \over dt}{\phi}_{2}^{i}(t) &=-\phi_{2}^{k}(t) a_{2,k}^{i}(t)\\
\non
{d \over dt}{\phi}_{2}^{ij}(t)&=-\phi_{2}^{k}(t)a_{1,k}^{ij}(t)-\phi_{2}^{kj}(t)b_{k}^{i}(t)-\phi_{2}^{ik}(t)b^{j}_{k}(t)\\
\label{eq:RECC1} 
{d \over dt}{\phi_1^{i}}(t)&=-\phi_1^{k}(t) b^{i}_{k}(t)\\ 
\non
{d \over dt}{c_1}(t)&={d \over dt}{c_2}(t)=0.
\end{align}
With the initial condition,
\beq
\f_2^i(t_e)= \d_1^i, \f_2^{ij}(t_e)=0, c_2(t_e)=-(10^3\GEV)^2
\eeq
and
\beq
\f_1^i(t_e)=\d_1^i, c_1(t_e)=-10^3\GEV,
\eeq
taken from \Eq{appcon},
we obtain the mapped condition functions as the solutions of \Eq{RECC1}.

Notice that a mapped condition function which equals to a dimensionful parameter at $t_e$, as 
\eqn{\f(t_e)=m_i^2 ~ {\rm or}~  \f(t_e)=M_i, \laq{completeset}} is a solution of the RG equation with an explicit 
form in terms of the dimensionful parameters at $t$. These solutions actually can be obtained by solving 
\Eq{RECC1} with proper initial conditions imposed where $c_1(t)=c_2(t)=0$.
Finally, substituting all the solutions corresponding to the parameters at $t_e$, we can map any condition functions to the parameter space at an arbitrary scale $t$.

\providecommand{\href}[2]{#2}\begingroup\raggedright\endgroup

\end{document}